\pdfoutput=1
\documentclass[namedreferences]{solarphysics}
\usepackage[hyperref,optionalrh,solaromanenum]{spr-sola-addons} 
\usepackage{graphicx}                    
\usepackage{color}                       
\usepackage{breakurl}                         

\begin{document}
\begin{article}
\begin{opening}

\title{Transition of the Sunspot Number from Zurich to Brussels in 1980: A Personal Perspective}

\author[addressref={1,2},corref,email={stenflo@astro.phys.ethz.ch}]{\inits{J.O.}\fnm{J.O.}~\lnm{Stenflo}}

\runningauthor{J.O. Stenflo}
\runningtitle{Transition of the Sunspot Number from Zurich to Brussels}

\address[id={1}]{Institute for Astronomy, ETH Zurich, CH-8093 Zurich, Switzerland}
\address[id={2}]{Istituto Ricerche Solari Locarno, Via Patocchi, CH-6605 Locarno-Monti, Switzerland}

\begin{abstract}
The Swiss Federal Observatory, which had been founded in 1863 by Rudolf Wolf,
was dissolved in connection with the retirement of Max Waldmeier in
1979. The determination of the Zurich sunpot number, which had been a
cornerstone activity of the observatory, was then discontinued by ETH 
Zurich. A smooth transition of the responsibility for the sunspot
number from Zurich to Brussels could however be achieved in 1980,
through which it 
was possible to avoid a discontinuity in this important time
series. Here we describe the circumstances that led to the termination
in Zurich, how Brussels was chosen for the succession, and how the
transfer was accomplished. 
\end{abstract}

\keywords{Sunspots, Statistics; Solar Cycle, Observations}

\end{opening}

\section{Historical Overview}\label{sec:hist} 
\begin{figure} 
\centerline{\includegraphics[width=0.75\textwidth,clip=]{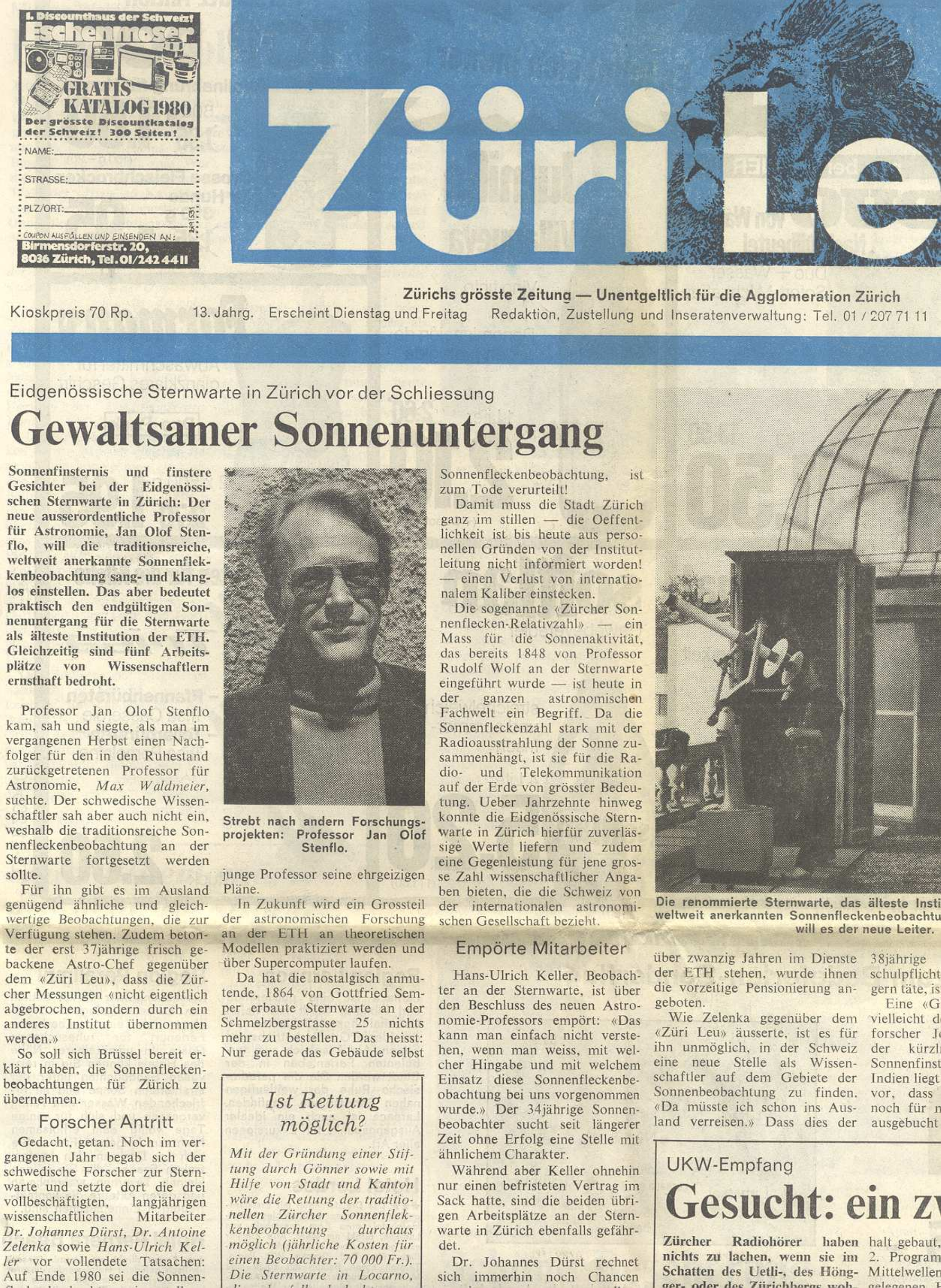}}
\caption{Front-page news 20 May 1980:  ``Violent sunset'' at the Swiss
  Federal Observatory.}\label{fig:zuerileu} 
\end{figure}

Although it is well known that the leading role that Zurich had in
establishing and determining the relative sunspot number ended with
Max Waldmeier's  retirement from ETH Zurich, and that the responsibility for the
continuation of this program was subsequently taken over by Brussels, few people
really know why and how all this happened. As the successor on the ETH 
Chair of Max Waldmeier I happened to play a key role in this
transition. In the popular media I was initially blamed for coming
to Switzerland as a young Swede and without sensitivity abruptly
discontinuing a century-old Swiss tradition. Figure~\ref{fig:zuerileu}
shows parts of the front page of a paper (Z\"uri Leu) that was distributed free
of charge to all households in the Zurich area. 

\begin{figure} 
\centerline{\includegraphics[width=0.9\textwidth,clip=]{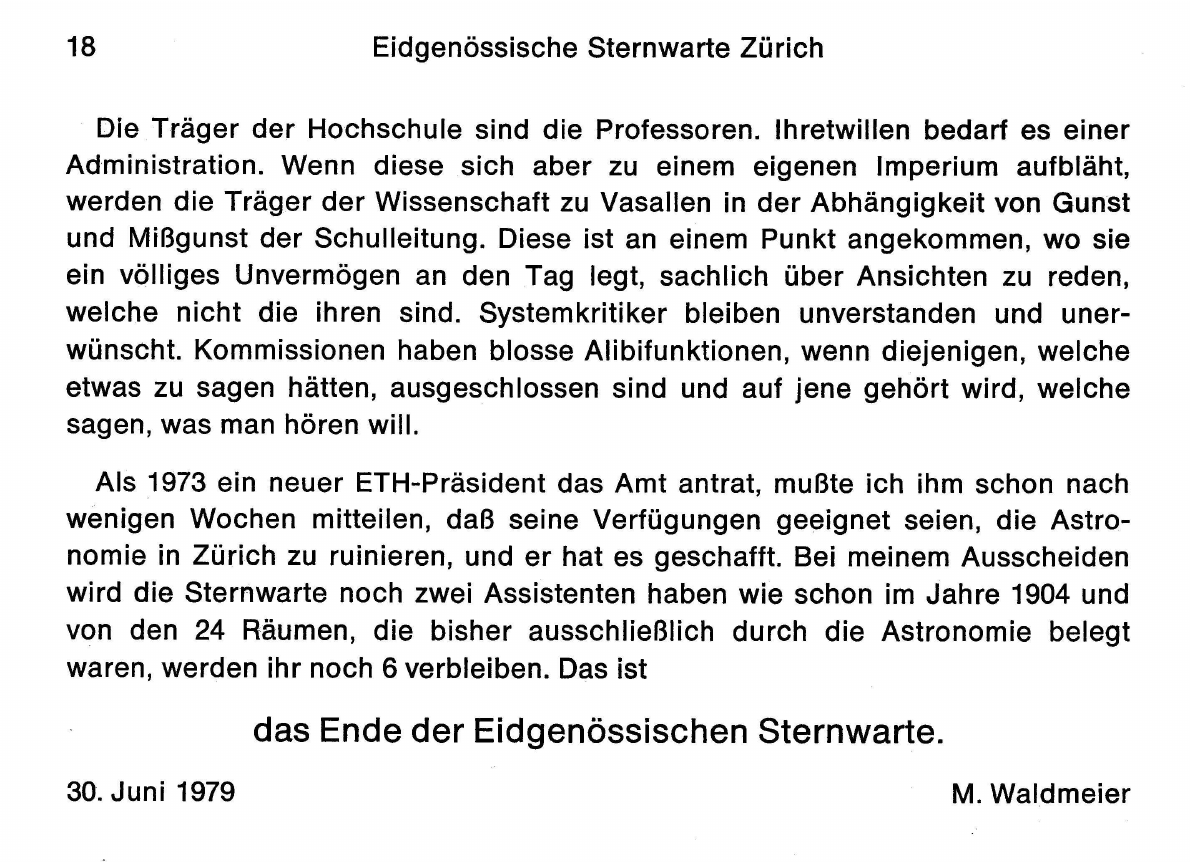}}
\caption{Last page from Waldmeier's final Annual Report, signed 30 June
  1979, before his successor had been chosen. He declares that the
  ``end of the Swiss Federal Observatory'' has happened through a
  process that started many years before. }\label{fig:wald79}
\end{figure}

With his strong personality Max Waldmeier had over the years made  
many enemies within the ETH, something that I learnt when I came there,
although my own encounters with him had always been very friendly and
even cordial. Figure~\ref{fig:wald79} shows how Waldmeier in 1979
ended his last Annual Report \citep{stenflo-waldmeier79} before his
successor had been chosen, 
bitterly accusing the ETH of pursuing a policy over the previous several
years with the aim of ``ruining astronomy in Zurich''. The ETH wanted
to use the opportunity of Waldmeier's 
retirement to dissolve the Swiss Federal Observatory (Eidgen\"ossische
Sternwarte) to make a fresh start, preferrably in an entirely new field of 
astronomy. Thus a US extragalactic astronomer was first offered the
position, but as the negotiations with him fell through, 
the search committee had to make a new
choice, and this time I got the offer. 
So it happened that solar physics could after all be
continued in Zurich. 

ETH dissolved the Eidgen\"ossische Sternwarte and all of the positions
related to the sunspot-number program a few months before I began my
position on 1 April 1980, to give me a ``clean table'' to start
with. However, I naturally share responsibility for what happened,
since ETH asked for my consent as the new professor before they
executed these decisions. It was fully clear to me that the
sunspot-number time series was of great importance to the wide
scientific community and had to be continued. So why did I give my consent to
this\,?  

There were two main reasons: i) I saw that satisfactory alternative
solutions for a responsible continuation of the sunspot-number time
series existed. The overriding priority was that reliable
determinations could be continued somewhere. For me with an
international outlook it was not essential that it had to continue to
be in Zurich. ii) Once I had come to understand the atmosphere at the ETH,
it was clear that stubbornly clinging to the continuation of the
Zurich role would block the opportunities to develop new research
programs in solar physics. 

Nevertheless it was also clear to me that the identification of a long-term
solution to secure the continuation of the sunspot number had to be my
first priority at the ETH, and that swift action on this matter was
essential. In anticipation of Waldmeier's retirement there had been
discussions within the IAU about the long-term future 
of the sunspot number. In this context Alan Shapley of NOAA in
Boulder, Colorado, USA, had done detailed statistical studies \citep{stenflo-shapley79} that showed 
excellent correlations between the Zurich sunspot number on the one
hand and the corresponding number determined by the American
Association of Variable Star Observers (AAVSO) on the other hand, as
well as with the solar 10-cm radio flux measured at Ottawa. A large
number of different observing stations sent their observations to
Zurich, where the various inputs were weighted together to form the
relative sunspot number. 

Among the potential candidates, who were interested in taking over the
responsibility for the 
determination of the sunspot number, were Madrid, Istanbul, Manila, and
Pulkovo (Leningrad), but without any particular evaluation procedure
it was clear to me that the best and most reliable choice would be the 
Observatoire de Bruxelles, where they had excellent experience over
many years as a contributing station to the Zurich sunspot number. It
was of great value that I happened to have been personally acquainted with the
Director, Andr\'e\  Koeckelenbergh, for several years, and I knew that
he was highly 
motivated and conscientious, perfectly suited for taking over this
task in a most responsible way. He had participated in a Workshop on
Solar Polarization that I organized in Lund, Sweden, in 1977, so we
had common scientific interests. 

Therefore, only two weeks after I immigrated to Switzerland from
Sweden and started my new job in Zurich I traveled to Brussels in
mid-April 1980 to have direct discussions with Koeckelenbergh about
the possibility of transferring the responsibility for the sunspot number
from Zurich to Brussels and about the modalities for this
transfer. From his enthusiastic response it was clear to me that this
would be an excellent solution that should be implemented as soon as
possible. During this process I also learnt that the most important
observing station
that carried by far the greatest weight in the determinations of the
relative sunspot number was Specola Solare in Locarno, which had
belonged to the Eidgen\"ossische Sternwarte under Waldmeier, but which
was taken over from the ETH by a local private foundation,
Associazione Specolar Solare Ticinese (ASST), when ETH discontinued
the sunspot number program. The long-term sunspot
observer, Sergio Cortesi, could continue his work there, but now
effectively in the capacity of being the {\it de facto} Director of Specola. 

As a follow-up of my Brussels visit I organized a meeting on 4 June
1980, at the newly formed Institute of Astronomy of ETH Zurich that I 
was now directing. At this meeting we formulated a detailed plan,
based on a draft by Koeckelenbergh,  which described how the transfer
to Brussels should be executed, and which also defined the future role
that Specola should play in this context. This plan was unanimously
accepted by the participants at the meeting, who included besides 
Koeckelenbergh and Cortesi also Max Waldmeier and the two main 
representatives of the Zurich sunspot program, A. Zelenka and H.U. 
Keller. According to our plan the transfer would be complete by the
end of 1980, and from then on the designation would change from the
Zurich relative sunspot number to the International Sunspot
Number. Specola Solare with Sergio Cortesi as observer would continue
to serve as the main observing station and be given the same dominant  
weight also for the International Sunspot Number. This would assure
continuity so that the transition to Brussels would not lead to any 
glitch in the sunspot time series. 

Two days after our ETH meeting, on 6 June, I informed the President of IAU
Commission 10, Vaclav Bumba from Prague, about our plans. The IAU was
pleased with this solution and happy to support it. 

\begin{figure} 
\centerline{\includegraphics[width=0.5\textwidth,clip=]{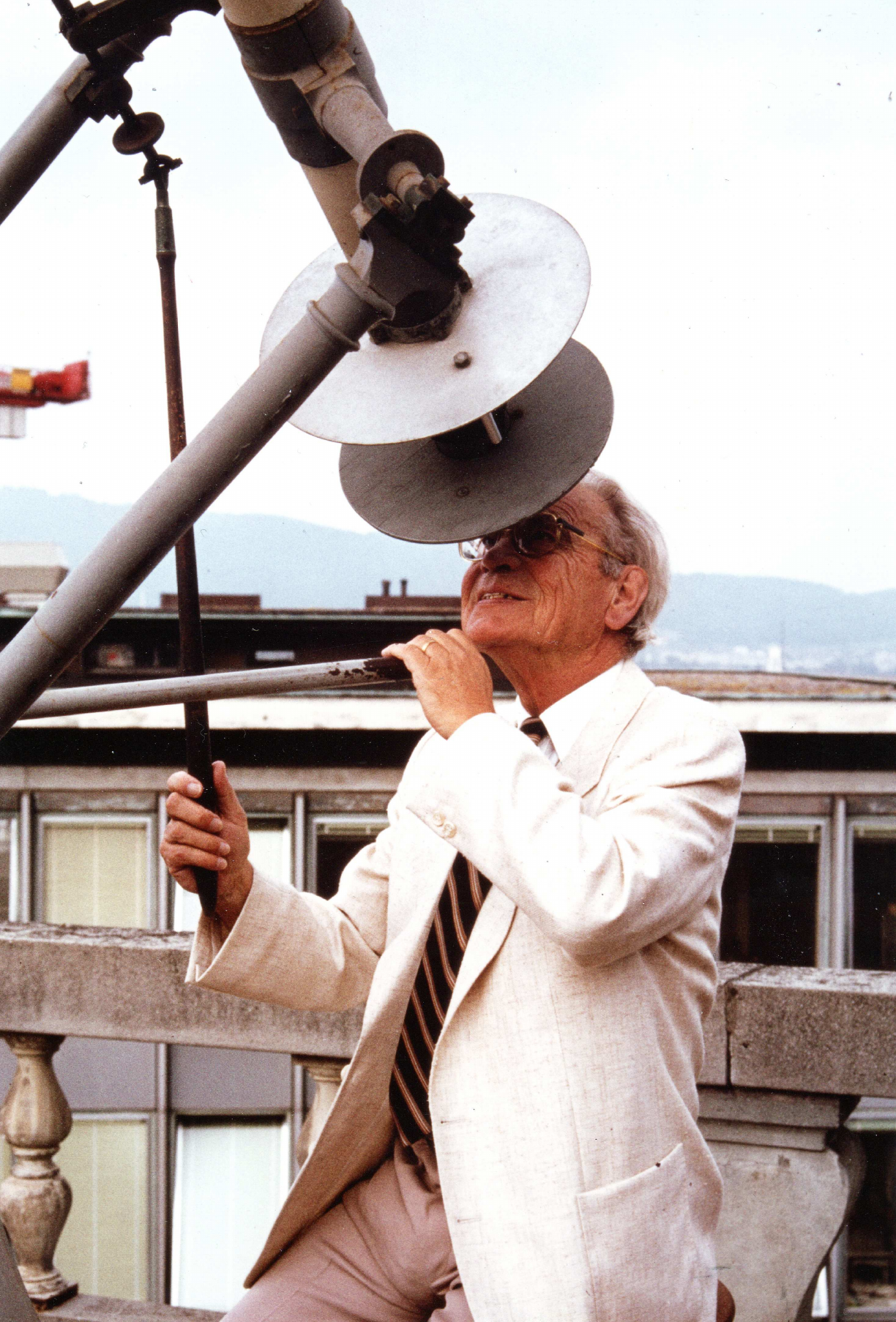}}
\caption{Max Waldmeier in 1983 looking through the Fraunhofer
  refractor that had been used since the time of Rudolf Wolf to count
  the sunspots.}\label{fig:wald83}
\end{figure}

The backing by IAU was essential to make it unambiguously clear
who was now in charge of the sunspot-number task, since there were
other organizations who kept publishing their versions of the sunspot
number. One of them was the Swiss military, who took over the funding
of the long-term observer at the Eidgen\"ossische
Sternwarte in Zurich (H.U. Keller) when ETH terminated his employment,
so that he could continue to count the spots as before. 
Figure~\ref{fig:wald83} shows Waldmeier in 1983 looking through the
Fraunhofer refractor, which continued to be used by H.U. Keller for
the determination of the sunspot number at the Zurich Sternwarte. 

\begin{figure} 
\centerline{\includegraphics[width=0.6\textwidth,clip=]{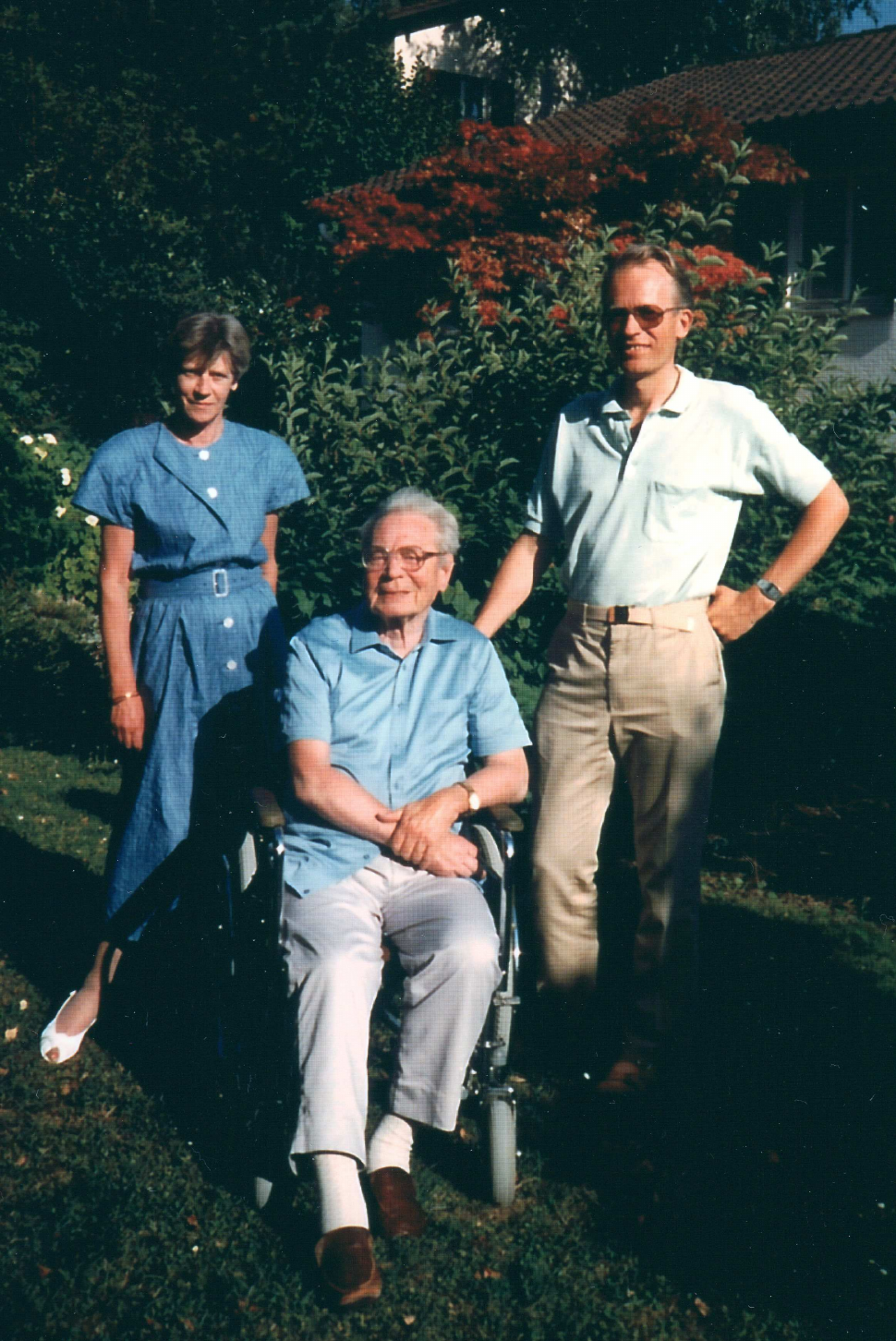}}
\caption{Max Waldmeier in the garden of his house in Zurich on 4 June 1992, during a
  visit by myself and Susi Weber, who served her whole professional life as
  astronomy secretary, first under Waldmeier, then for nearly two
  decades as my secretary.}\label{fig:wald92}
\end{figure}

\section{Epilogue}\label{sec:epi} 
 After having successfully served as a kind of ``midwife'' to secure the
long-term future continuation of the sunspot-number series through the
transfer from Zurich to Brussels, I consciously tried to keep a
distance from all of the activities related to the sunspot number, not
only from the militarily supported work in Zurich, but to some
extent also from the work at Specola, to avoid confusion, because
Brussels had to be viewed as the sole organization in charge, and it
needed to be clear that the
activities in my ETH institute dealt with other aspects of solar
physics. The distance that I kept from the sunspot number might have
been interpreted as a disinterest, but I have always appreciated the
importance and necessity of the sunspot record and have from time to
time myself been a scientific user of this record. 

 Waldmeier suffered a stroke in 1986, which paralyzed him and left him
in a debilitated state with little capacity left for speaking or
comprehending language. This sad state lasted until his death in
2000. Figure~\ref{fig:wald92} shows him in 1992 in his wheelchair
during a visit by myself and my secretary, Susi Weber, who had also
served for many years as Waldmeier's secretary. 

Looking back over the more than three decades that have since passed,
I feel vindicated that we came up with a good solution back in
1980. The sunspot number series is in good hands.



\end{article} 
\end{document}